\documentstyle[11pt,newpasp,psfig,twoside]{article}
\index{instructions}
\index{guidelines}

\def\edcomment#1{\iffalse\marginpar{\raggedright\sl#1\/}\else\relax\fi}
\reversemarginpar

\begin{document}
\title{Heavy Neutron Stars?  
A Status Report on Arecibo
Timing of Four Pulsar--White Dwarf Systems}

\author{David J. Nice \& Eric M. Splaver}
\affil{Physics Department, Princeton University\\ Box 708,
Princeton, NJ 08544 USA}
\author{Ingrid H. Stairs}
\affil{Department of Physics and Astronomy, University of British Columbia\\
6224 Agricultural Road, Vancouver, BC V6T 1Z1, Canada}

\begin{abstract}

Relativistic phenomena (orbital precession, Shapiro delay,
and/or orbital decay) have been measured in Arecibo timing
observations of four pulsar--white dwarf binaries, leading
to constraints on the neutron star masses.  We have detected
the decay of the PSR J0751+1807 orbit due to gravitational
radiation emission, the first such measurement in a binary
with a low mass ratio ($m_2/m_1\ll 1$).  Timing data
constrains the mass of this pulsar to bet between 1.6 and
2.8\,M$_{\sun}$.  Masses of the other pulsars are in
marginal agreement with the canonical pulsar mass of
1.35\,M$_{\sun}$, but higher values are preferred.

\end{abstract}

\section{Introduction}

The conventional wisdom is that all pulsars have mass close
to the Chandrasekhar value, $\sim 1.35\,{\rm M}_{\sun}$.  This belief has
arisen from high precision measurements of 
neutron star--neutron star binary systems;  in all cases,
the pulsar and the secondary
neutron star have masses within a narrow range, 1.33 to 1.45\,M$_{\sun}$
(Stairs et al. 2002, Weisberg \& Taylor 2003,
Deich \& Kulkarni 1996).  
Perhaps all neutron stars are born
within this narrow mass range, or perhaps
the particular evolution of double neutron star
leads to this result (Brown 1995).
Even if all neutron stars are born near the
Chandrasekhar mass, they may achieve
higher masses through accretion of matter from
orbital companions.  Millisecond pulsars in orbits
with white dwarfs, 
in particular, are known to have accreted
mass during their low-mass X-ray binary phases,
and hence they are
prime candidates to be relatively massive
pulsars.  There has been some observational
evidence for heavy pulsars in these systems
(Kaspi, Taylor, \& Ryba 1994; van Straten et al. 2001),
although a statistical study argued that they are no
heavier than pulsars in double neutron star binaries
(Thorsett \& Chakrabarty 1999).

In this status report, we summarize observations
of four millisecond pulsars in pulsar--white
dwarf binaries
made with the Arecibo telescope.
This is an update of
Nice, Splaver, \& Stairs (2003), incorporating 
significant new
timing data from 
PSR~J0751+1807 and an improved timing model
(including orbital-annual parallax) for PSR J1713+0747.

\begin{table}[t]
\caption{Mass Measurements from Timing Analysis}
\begin{tabular}{l@{}lcccc}
\tableline
\tableline
\multicolumn{2}{c}{Pulsar} & Orbital  & Eccentricity & Mass         & Pulsar Mass \\
             &             & Period   &              &  Function    &  (95\% confidence) \\
             &             & $P_b$ (days)   &   $e$  & $f_1$ (M$_{\sun}$) & $m_1$ (M$_{\sun}$) \\
\tableline
J&0621+1002 & \phantom{0}8.32 & 0.002\,457 & 0.0270 & 1.1$-$2.3 \\
J&0751+1807 & \phantom{0}0.26 & 0.000\,003 & 0.0010 & 1.6$-$2.8 \\
J&1713+0747 &           67.83 & 0.000\,075 & 0.0079 & 1.2$-$2.1 \\
B&1855+09   &           12.33 & 0.000\,022 & 0.0056 & 1.4$-$1.8 \\
\tableline
\tableline
\end{tabular}
\end{table}

\section{Observations \& Data Reduction}

Observations were made at Arecibo over several 
years at radio frequencies 430 and 1410\,MHz, 
primarily using the Princeton Mark\,IV data
acquisition system (Stairs et al. 2000).  Details will be
given elsewhere.  These data were augmented by observations
at Green Bank (140 Foot telescope), Jodrell Bank, and
Effelsberg, as well as data collected at Arecibo using the
Arecibo Berkeley Pulsar Processor and the Princeton
Mark\,III system.

Pulse arrival times were derived using
conventional methods and were fit to timing models
using {\sc tempo}\footnote{http://pulsar.princeton.edu/tempo}.
For each pulsar binary, at least one post-Keplerian orbital
parameter is detectable, leading to constraints on pulsar mass,
$m_1$, secondary star mass, $m_2$, and orbital 
inclination, $i$. (See Taylor, 1992, for the relations between
post-Keplerian parameters, masses, and orbital inclination.)
To derive confidence intervals on $m_1$, $m_2$, and $i$, we 
calculated the goodness-of-fit of
timing solutions over a grid of trial values of $\cos i$ and $m_2$.
Each pair of trial values corresponds to a particular pulsar mass,
$m_1$, according to the mass function,
$
f_1=(m_2\sin i)^3(m_1+m_2)^{-2}
  =(2\pi/P_b)^2x^3/T_{\sun},
$
where $T_{\sun}=4.925\,\mu$s is the solar mass in time units,
$P_b$ is the orbital period, $x$ is the semi-major axis projected
into the line-of-sight in time units, and $m_1$ and $m_2$ are in solar mass units.
For each trial, relativistic 
post-Keplerian parameters appropriate to 
$\cos i$ and $m_2$ were calculated and held constant, while all
other parameters were fit.

The contours plotted in Figure 1 
show the combinations of $\cos i$ and $m_1$ that gave acceptable
fits for each source.  We expect $\cos i$ to be uniformly distributed
for randomly oriented binary systems.  (Other than this, the
intrinsic values of $i$ are of limited interest.)
Table 1 summarizes the pulsar parameters and 95\% confidence
limits on pulsar masses from the timing data.

\begin{figure}[t]
\centerline{\psfig{file=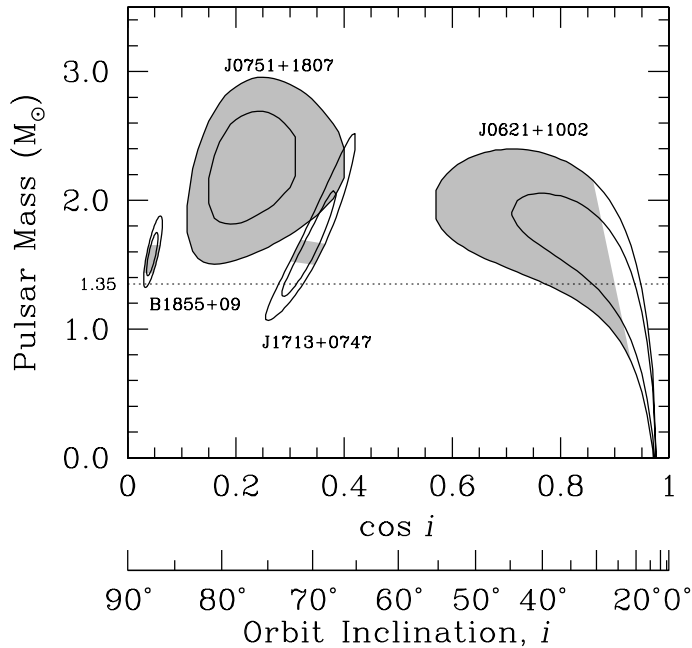,width=0.7\textwidth}}
\caption{Constraints on pulsar masses and
orbital inclinations.  Inner and
outer contours are 68\% and 95\% confidence
regions, respectively, based on timing data alone.
Shaded areas indicate theoretical limits within these
contours, from the assumption that the companion
is a white dwarf (J0621+1002) and from predictions
of secondary star masses from the $P_b-m_2$ relation (B1855+09
and J1713+0747).  (There are no theory constraints
on J0751+1807.)}
\end{figure}

\section{Individual Sources}

\underline{\it PSR J0621+1002}.  This is an ``intermediate mass'' 
binary millisecond pulsar, presumably with a CO white dwarf
secondary.  Constraints on the masses come from the
relativistic advance of periastron---we measure the
precession rate to be
$\dot{\omega}=0.0116\pm 0.0008^\circ\,{\rm yr}^{-1}$
($1\sigma$), yielding $m_1+m_2=2.8\pm0.3\,{\rm M}_{\odot}$---and
from the lack of detection of Shapiro delay, which excludes
large inclinations.  While the canonical pulsar mass value 
of 1.35\,M$_{\sun}$ falls within the allowed region
(Fig. 1), higher values are preferred, especially
when we include the constraint that the white dwarf
secondary have mass less than 1.4\,M$_{\sun}$ (shading
in Fig. 1).  See Splaver et al. (2002) for more details.

\underline{\it PSR J0751+1807}.  This pulsar has a very short orbital
period.  By combining pre-upgrade (1993--1994) and post-upgrade
(1999--2003) Arecibo observations of this source, along with
data from Jodrell Bank and Effelsberg, we are able to detect
the orbital decay due to emission of gravitational
radiation.  
The rate of change of orbital period is 
$\dot{P_b}=(-6.2\pm1.1)\times 10^{-14}$.  This is the first 
detection of the decay of a pulsar--white dwarf 
orbit with a low mass ratio 
($m_2/m_1\ll 1$).\footnote{Note also the spectacular results from 
PSR J1141$-$6545, a young pulsar in an eccentric orbit with a massive
white dwarf
(Bailes et al. 2003.)}  Shapiro delay is marginally
detected, constraining $i$ to intermediate values.
If the Shapiro delay detection can be confirmed and refined
it will provide independent constraints on
the system masses, allowing for novel tests of relativistic
gravity (Arzoumanian 2003).  
In any case, the allowed values of $m_1$ are significantly greater
than 1.35\,$M_{\sun}$  (Fig.~1).  

\underline{\it PSRs B1855+09 \& J1713+0747}.
These pulsars are in wide orbits with light white dwarf
companions.  In both cases, the Shapiro delay is detectable,
constraining $i$ and $m_1$ as shown in Figure 1.
Binary evolution theory posits
a narrow range of companion masses, $m_2$, for a given orbital period, $P_b$,
in these systems
(Podsiadlowski, Rappaport, \& Pfahl 2002; Tauris \& Savonije 1999; 
Rappaport et al. 1995).  The resulting constraints on $m_1$ and
$i$ are indicated in Figure 1.  Once again, values of $m_1$ 
somewhat higher than 1.35\,M$_{\sun}$ are 
preferred.\footnote{ Parkes
pulsar J0437-4715 gives a similar result, $m_1=1.58\pm0.18\,{\rm M}_{\sun}$
(van Straten et al. 2001).}

\section{Conclusion}

The sources individually provide some evidence for
pulsars substantially more massive than
the Chandrasekhar value, and the collection of constraints
shown in Figure 1 provides a compelling case for heavy neutron stars.

\acknowledgments

The Arecibo Observatory is operated by Cornell University under a
cooperative agreement with the NSF.  Pulsar
research at Princeton University is supported by NSF grant 0206205.
IHS is supported by an NSERC UFA and Discovery Grant.  We thank
numerous collaborators, particularly Jim Cordes,
who provided crucial early data on PSR J0751+1807.


\begin{references}

\reference Arzoumanian, Z. 2003,
ASP Conf. Ser. 302, 69
\reference Bailes, M, Ord, S. M., Knight, H. S., \& Hotan, A W.
2003, \apj, 595, 49
\reference Brown, G. E. 1995, \apj, 440, 270
\reference Deich, W. T. S., \& Kulkarni, S. R. 1996,
Compact Stars in Binaries: IAU Symp. 165,
J. van Paradijs, E. P. J. van den Heuvel, \& E. Kuulkers,
eds. (Dordrecht: Kluwer), 279.
\reference Nice, D. J., Splaver, E. M., \& Stairs, I. H.
2003, ASP Conf. Ser. 302, 75
\reference Kaspi, V. M., Taylor, J. H., \& Ryba, M. F.
1994, \apj, 428, 713
\reference Podsiadlowski, P., Rappaport, S., \& Pfahl, E. D. 2002 \apj, 565, 1107
\reference Rappaport, S., et al.
1995,  \mnras, 273, 731
\reference Splaver, E. M.,  et al.  2002, \apj, 581, 509
\reference Stairs, I. H., Thorsett, S. E., Taylor, J. H., \& Wolszczan, A.
2002, \apj, 581, 501
\reference Tauris, T.~M., \& Savonije, G.~J. 1999, \aap, 350, 928
\reference Taylor, J. H. 1992, Phil. Trans. R. Soc. Lond. A, 341, 117
\reference Thorsett, S. E., \& Chakrabarty, D.
1999, \apj, 512, 288
\reference Weisberg, J. M., \& Taylor, J. H.  
2003, ASP Conf. Ser. 302, 93
\reference van Straten, W., et al.
2001,  Nature, 412, 158



\end{references}
\end{document}